\documentclass[conference]{IEEEtran}
\IEEEoverridecommandlockouts
\usepackage{cite}
\usepackage{amsmath,amssymb,amsfonts}
\usepackage{algorithmic}
\usepackage{graphicx}
\usepackage{textcomp}
\usepackage{xcolor}
\usepackage{graphicx}
\usepackage{subfigure}
\usepackage{amssymb,amsmath}
\usepackage{textcomp}
\usepackage{cite}
\usepackage{float}
\usepackage{bm}
\usepackage{here}
\usepackage{soul, color}
\usepackage{multirow}
\usepackage{booktabs}
\usepackage{multicol}
\usepackage{verbatim}
\usepackage{svg} 
\usepackage{graphicx}
\usepackage{CJKutf8}

\usepackage{amsmath}
\usepackage{amssymb,amsmath}
\usepackage{textcomp}
\usepackage{cite}
\usepackage{float}
\usepackage{bm}
\usepackage{here}
\usepackage{soul, color}
\usepackage{multirow}
\usepackage{booktabs}
\usepackage{multicol} 
\usepackage{array}

\usepackage{graphicx}
\usepackage{caption}

\def\BibTeX{{\rm B\kern-.05em{\sc i\kern-.025em b}\kern-.08em
    T\kern-.1667em\lower.7ex\hbox{E}\kern-.125emX}}
\begin{document}
\begin{CJK}{UTF8}{gbsn}
\title{{An Effective Current Limiting Strategy to Enhance Transient Stability of Virtual Synchronous Generator}\\

\thanks{This work is financially supported by Chinese National Natural Science Foundation(52277192), Chinese National Natural Science Foundation(52277191)。}
}

\author{
    \IEEEauthorblockN{Yifan Zhao$^{1}$, Zhiqian Zhang$^{1}$, Ziyang Xu$^2$, Zhenbin Zhang$^{1*}$, Jose Rodriguez$^{3}$.}
    \IEEEauthorblockA{$^1$ School of Electrical Engineering, Shandong University, Jinan, China}
    \IEEEauthorblockA{$^2$ Department of Electrical Engineering, Shanghai Jiao Tong University, Shanghai, China}
       \IEEEauthorblockA{$^3$ Faculty of Engineering, Universidad San Sebastian Santiago, 8420524, Chile.}
    \IEEEauthorblockA{yifan.zhao@mail.sdu.edu.cn, zhiqian.zhang@mail.sdu.edu.cn, zbz@sdu.edu.cn, keyonxzy@sjtu.edu.cn, jose.rodriguezp@uss.cl}
}
\vspace{-2pt}

\maketitle 

\begin{abstract}
VSG control has emerged as a crucial technology for integrating renewable energy sources. However, renewable energy have limited tolerance to overcurrent, necessitating the implementation of current limiting (CL)strategies to mitigate the overcurrent. The introduction of different CL strategies can have varying impacts on the system. While previous studies have discussed the effects of different CL strategies on the system, but they lack intuitive and explicit explanations. Meanwhile, previous CL strategy have failed to effectively ensure the stability of the system.
In this paper, the Equal Proportional Area
Criterion (EPAC) method is employed to intuitively explain how different CL strategies affect transient stability. Based on this, an effective current limiting strategy is proposed. Simulations are conducted in MATLAB/Simulink to validate the proposed strategy. The simulation results demonstrate that, the proposed effective CL strategy exhibits superior stability.

\end{abstract}


\section{Introduction}
As the integration of Distributed Energy Resources (DERs) in power systems continues to grow, they present significant operational challenges due to their inherent weak damping and low inertia. These characteristics can lead to frequency fluctuations and voltage instability~\cite{xiaofan,currenttlintingsaeedifard,priealdll,deekpark2}, affecting the overall stability and reliability of the grid. To address these issues, VSG control strategy has been introduced~\cite{xiaona, zqz}. The VSG emulates the dynamics of traditional synchronous generators by mimicking the swing equations, thereby providing synthetic inertia and damping to enhance system stability. This strategy allows DERs to dynamically adjust their output in response to grid conditions, helping to maintain frequency and voltage within acceptable ranges. The VSG's flexibility and adaptability are key advantages, as it can quickly increase power output during frequency drops to support system recovery. Moreover, the VSG can be integrated with advanced control techniques such as model predictive control and optimization scheduling~\cite{xinliang,control}, further improving the efficiency and stability of the power system. The VSG control strategy offers an effective technical solution to the challenges posed by DERs, contributing to the sustainable and efficient operation of power systems.

During disturbances in VSG system, overcurrent may occur. However, devices have limited capability to withstand high current~\cite{current}, prompting the proposal of various current limiting strategies to mitigate overcurrent. Previous studies have introduced current limter, Virtual impedance and Voltage limter control to suppress overcurrent~\cite{review,impacts,liyapunuofu,currenttlintingsaeedifard}. For current limter control, there are typically three types: \textit{d}-axis priority, \textit{q}-axis priority and angle priority. The transient stability of a VSG is significantly affected by different current limiting strategies. It has been shown in other paper that taking the \textit{q}-axis current prioritized strategy exhibits improved transient stability performance. In~\cite{impacts}, it merely conducted experiments for validation without performing theoretical analysis. In~\cite{liyapunuofu}, it adopted Liapunov function method to analysis, but lacked intuitiveness. Although the \textit{q}-axis priority CL strategy has been verified to have better transient stability compared to other strategies, it still cannot guarantee the transient stability of the system under large disturbances.

In this paper, the Equal Proportional Area Criterion (EPAC) method is employed to intuitively and qualitatively explain how different current limiting strategies affect transient stability. Based on this analysis, an effective  current limiting strategy is proposed to effectively address transient stability issues encountered by VSG during disturbances. the simulation conducted in MATLAB/Simulink demonstrated that the proposed effective CL strategy enhances transient stability, thus offering a promising solution to improve the performance of VSG under large disturbances.
\section{VSG Control Principle And Previous Current Limiting Strategy}
This section describes the basic control strategy of VSG, then analyzes the existing current limiting strategies.
\subsection{VSG topology}

\begin{figure*}[htbp]
\begin{minipage}{0.32\linewidth}
		\vspace{3pt}
		\centerline{\includegraphics[width=\textwidth]{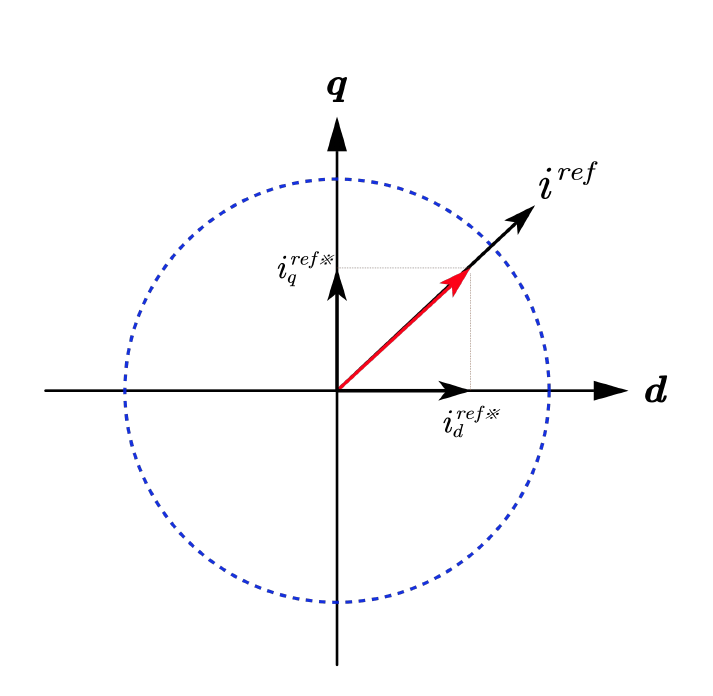}}
		\centerline{(a).}
	\end{minipage}
	\begin{minipage}{0.32\linewidth}
		\vspace{3pt}
		\centerline{\includegraphics[width=\textwidth]{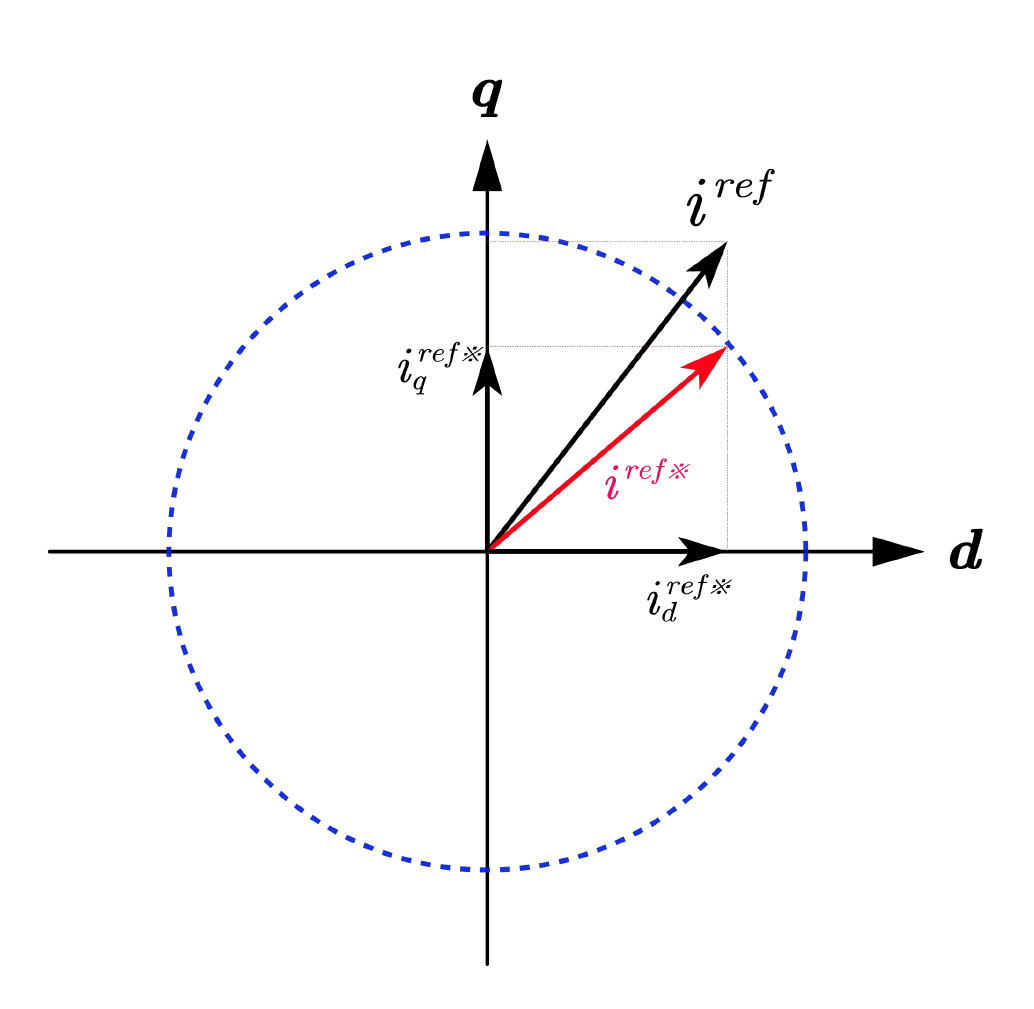}}
	 
		\centerline{(b).}
	\end{minipage}
	\begin{minipage}{0.32\linewidth}
		\vspace{3pt}
		\centerline{\includegraphics[width=\textwidth]{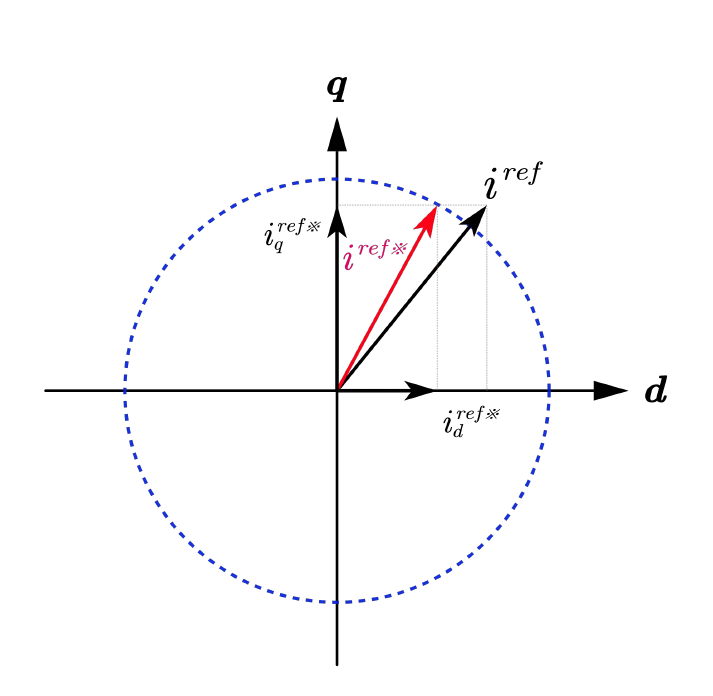}}
	 
		\centerline{(c).}
	\end{minipage}
 
	\caption{Three Current Limiter. (a) Angle priority CL. (b) \textit{d}-axis priority CL. (c) \textit{q}-axis priority CL.}
	\label{Fig_current limiting}
\end{figure*}

\begin{figure}[h]
	\centering
	\includegraphics[width=1\linewidth]{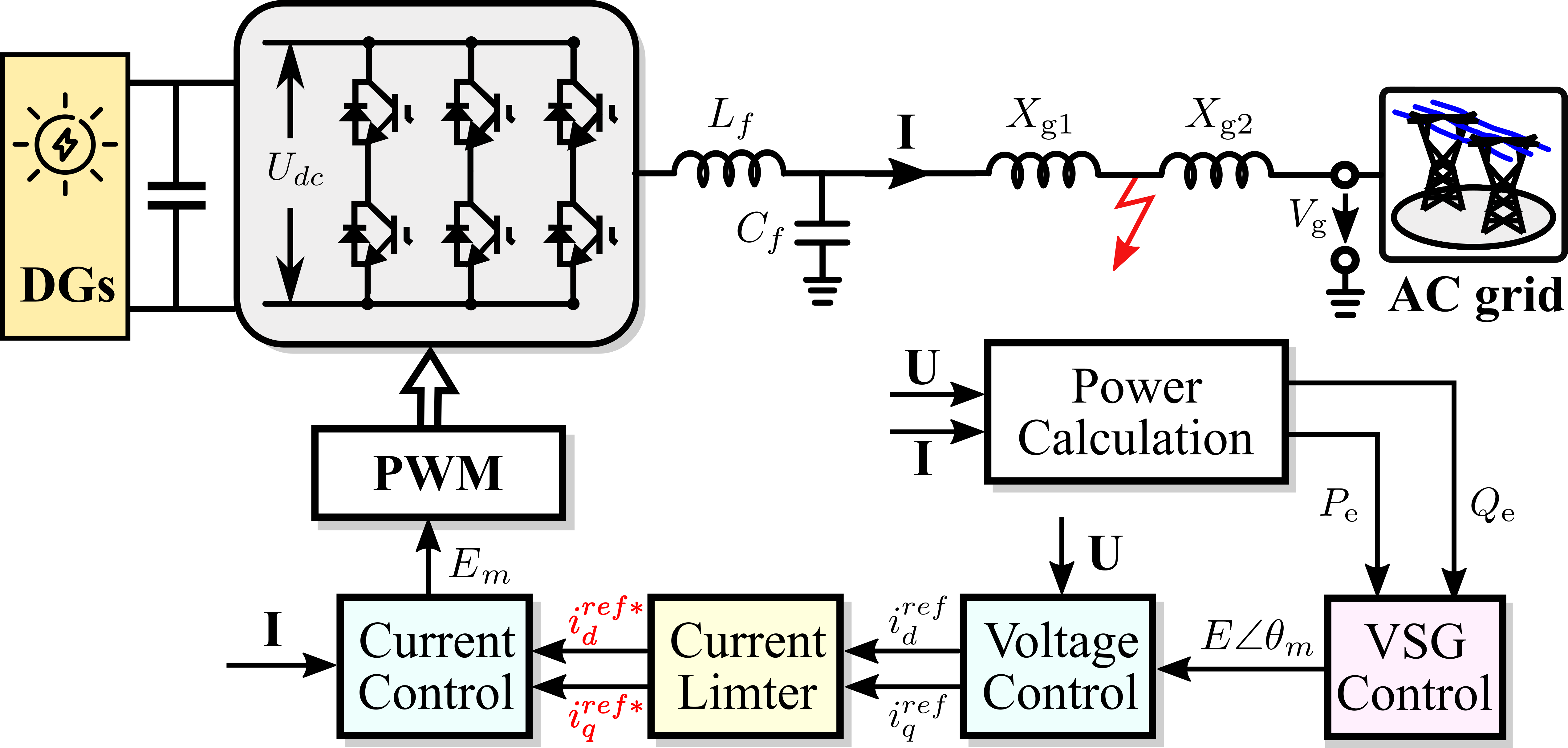}
    \caption{Topology and control scheme of VSG.}
	\label{Fig1 topology}
\end{figure}

The basic topology of the grid-tied VSG control strategy is given by Fig.~\ref{Fig1 topology}, \( U_{d c} \) is the dc-side voltage, \( I \) and
\( U \) are the VSG output current and voltage, and \( L_f \) is the filter
inductance. \( C_f \) is the filter capacitance, the \( I \) and \( U \) are the VSG
output current and voltage, \( L_g \) is the grid inductance and \( V_g \) is the grid voltage.

The VSG control loop consists of swing equation~\eqref{Equ:swing}and an automatic voltage regulator (AVR)~\eqref{Equ:avr}.
As shown in~\eqref{Equ:swing}, Swing equations are used to mimic the electromagnetic characteristics of synchronous generators. \( P_m \) and \( P_e \) are the active power reference and output active power. \( J \) is the virtual interia, \( D \)  is damping coefficient. \( w_m \) and \( w_0 \)  are virtual rotor frequency and nominal frequency. $\delta$ is the power angle.

\begin{equation}
	P_m-P_e=J \omega_m \frac{d \omega_m}{d l}+D\left(\omega_m-\omega_0\right)
 \label{Equ:swing}
\end{equation}

\begin{equation}
	E=E_{\mathrm{ref}}+k\left(Q_{\mathrm{ref}}-Q_e\right)
 \label{Equ:avr}
\end{equation}

\begin{equation}
	\delta=\int \omega_m-\omega_0 \mathrm{~d} t
 \label{Equ:theta}
\end{equation}

As shown in~\eqref{Equ:avr}. AVR is used to mimic characteristics of the voltage regulation in synchronous generators. \( E \) and $E_{\text {ref }}$ are the voltage rated and the amplitude of reference voltage voltage rated. \( k \) is the $Q-U$ coefficient value.

\subsection{Previous Current Limiting Strategy}

When the VSG experiences transient state, overcurrent occurs, necessitating the implementation of a CL, previous paper have proposed various current limiting strategy to reduce the current, including Current limter, Virtual impedance, Voltage limiter. In this paper, only current limiter is discussed.

\subsubsection{Angle Priority CL}

As shown in Fig.~\ref{Fig_current limiting}(a), The angle priority CL strategy ensure the power angle of $i^{r e f}$ remains unchanged after passing through the CL controller, $i_d^{\text {ref※ }}$ and $i_q^{\text {ref※ }}$ are the new reference current after CL. They are determined by~\eqref{equ:angle}.

\begin{equation}\label{equ:angle}
	\begin{aligned}
		& i_d^*=\frac{i_d^{\mathrm{ref}}}{\left|i_d^{\mathrm{ref}}\right|} \times \min \left(\left|i_d^{\mathrm{ref}}\right|,\left|i_d^{\mathrm{ref}}\right| \times \frac{I_{\mathrm{max}}}{\sqrt{\left(i_d^{\mathrm{ref}}\right)^2+\left(i_q^{\mathrm{ref}}\right)^2}}\right) \\
		& i_q^*=\frac{i_q^{\mathrm{ref}}}{\left|i_q^{\mathrm{ref}}\right|} \times \min \left(\left|i_q^{\mathrm{ref}}\right|,\left|i_q^{\mathrm{ref}}\right| \times \frac{I_{\max }}{\sqrt{\left(i_d^{\mathrm{ref}}\right)^2+\left(i_q^{\mathrm{ref}}\right)^2}}\right) \\
		&
	\end{aligned}
\end{equation}

\subsubsection{\textit{d}-axis priority CL}

As shown in Fig.~\ref{Fig_current limiting}(b), \textit{d}-axis priority CL keeps $i_d^{r e f}$  as constant as possible and achieves current limiting by cutting $i_q^{r e f}$, $i_d^{\text {ref※ }}$ and $i_q^{\text {ref※ }}$ are determined by~\eqref{equ:d}.

\begin{equation}
\label{equ:d}
	\begin{aligned}
		& i_d^*=\frac{i_d^{\mathrm{ref}}}{\left|i_d^{\mathrm{ref}}\right|} \times \min \left(\left|i_d^{\mathrm{ref}}\right|, I_{\mathrm{max}}\right) \\
		& i_q^*=\frac{i_q^{\mathrm{ref}}}{\left|i_q^{\mathrm{ref}}\right|} \times \min \left(\left|i_q^{\mathrm{ref}}\right|, \sqrt{\left(I_{\max }\right)^2-\left(i_d^*\right)^2}\right)
	\end{aligned}
\end{equation}
\begin{figure*}[htbp]
\begin{minipage}{0.32\linewidth}
		\vspace{3pt}
		\centerline{\includegraphics[width=\textwidth]{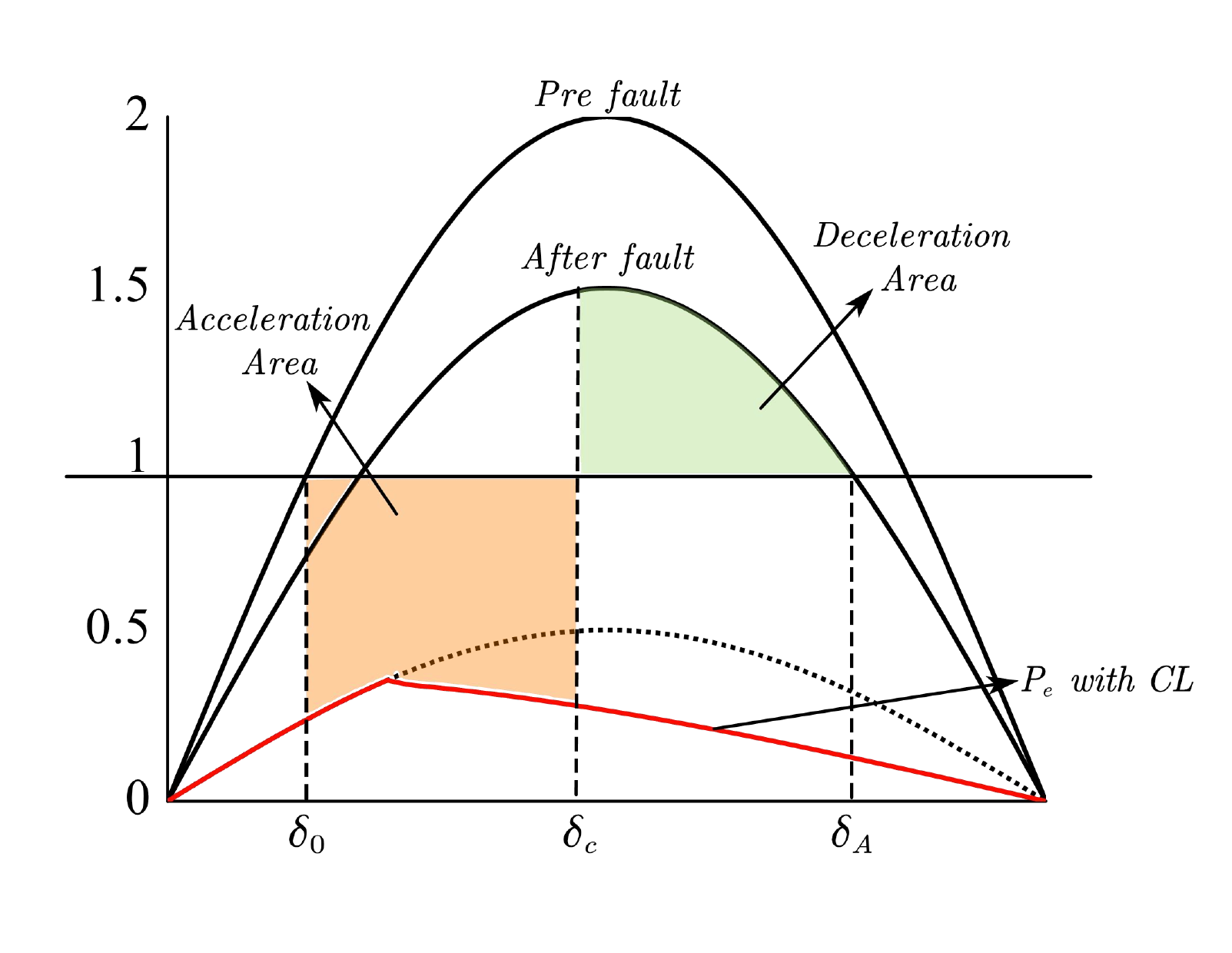}}
		\centerline{(a).}
	\end{minipage}
	\begin{minipage}{0.32\linewidth}
		\vspace{3pt}
		\centerline{\includegraphics[width=\textwidth]{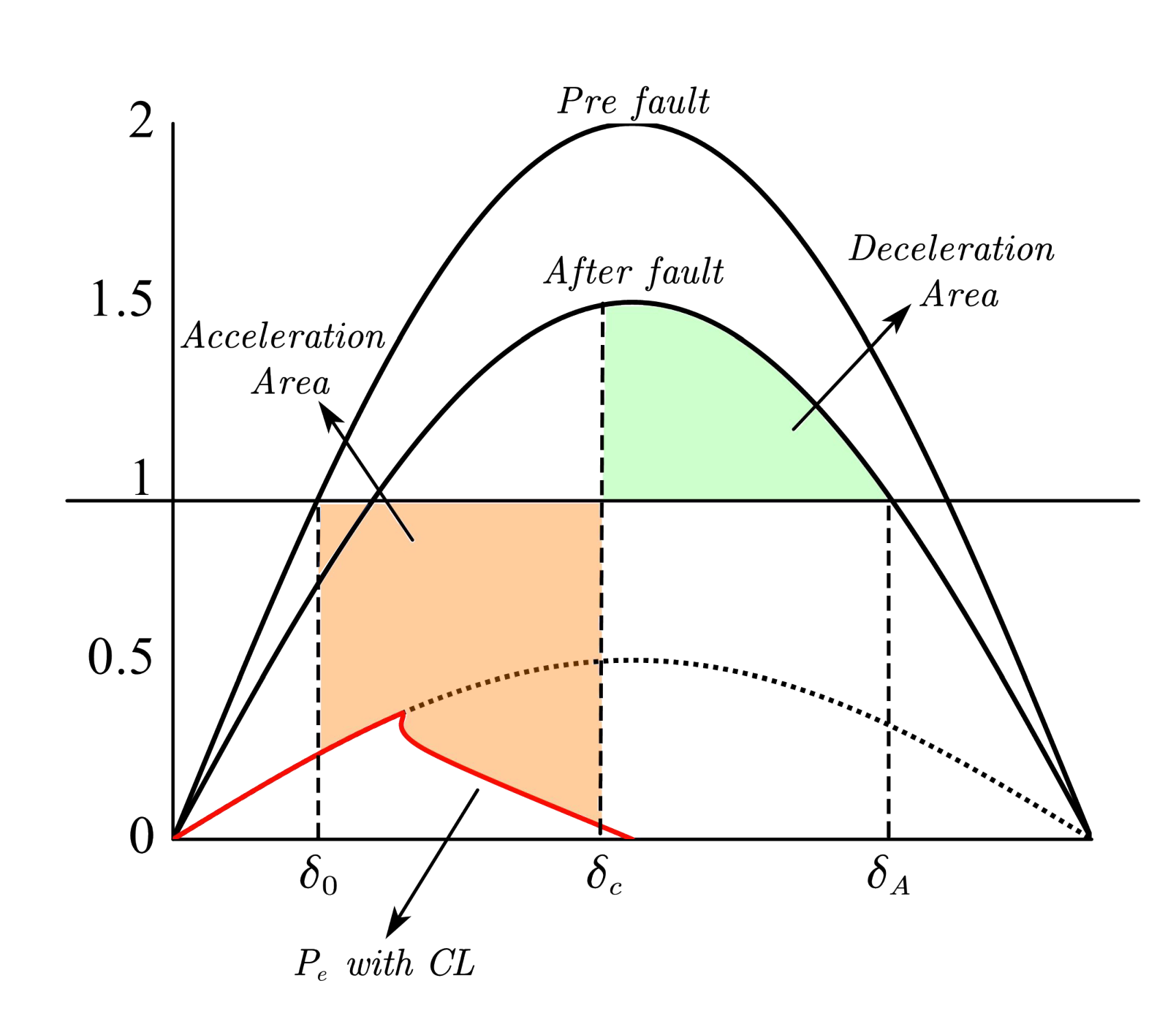}}
	 
		\centerline{(b).}
	\end{minipage}
	\begin{minipage}{0.32\linewidth}
		\vspace{3pt}
		\centerline{\includegraphics[width=\textwidth]{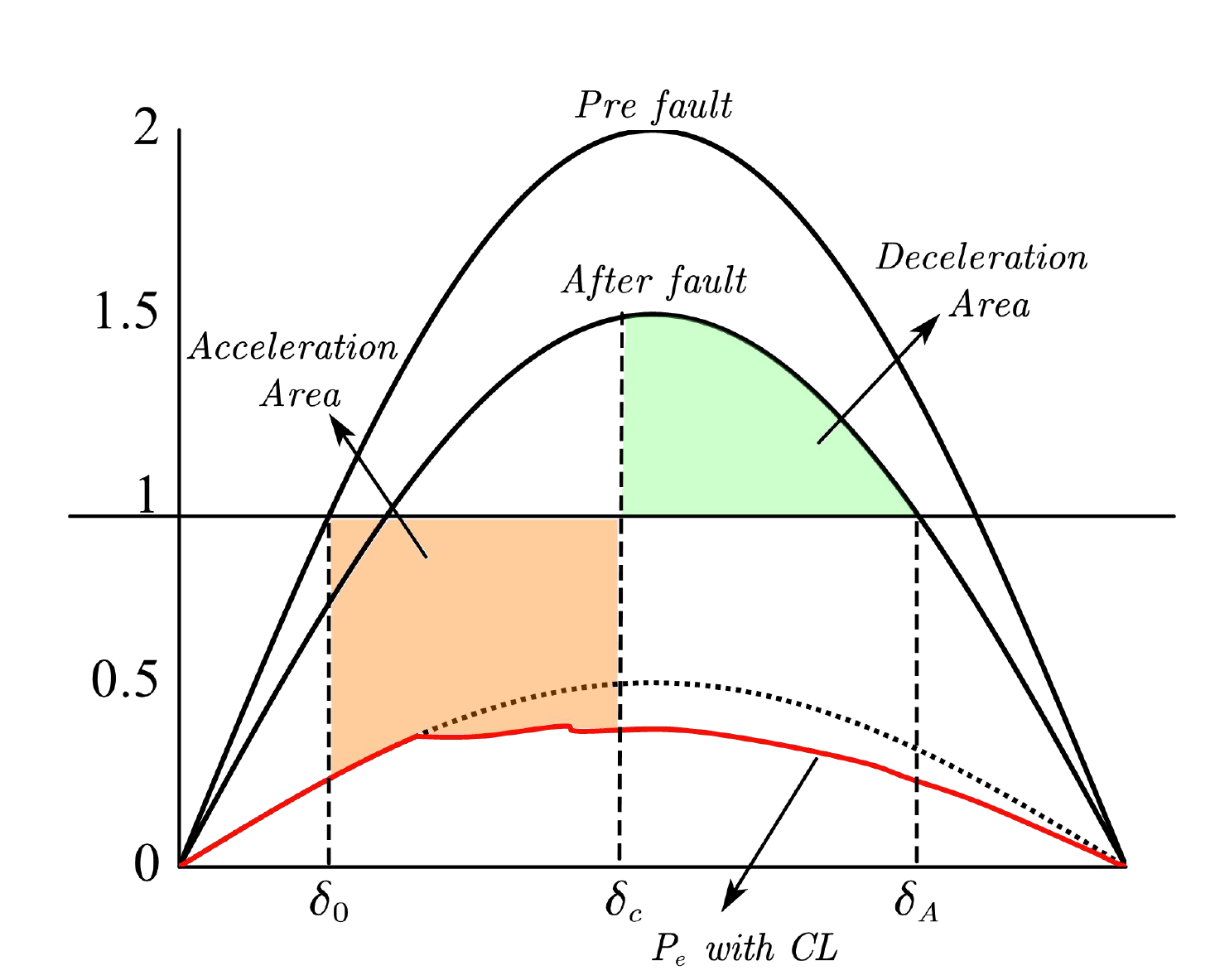}}
	 
		\centerline{(c).}
	\end{minipage}
 
	\caption{Transient stability analysis using EPAC method. (a) Angle priority CL. (b) \textit{d}-axis priority CL. (c) \textit{q}-axis priority CL.}
	\label{Fig_epac}
\end{figure*}
\subsubsection{\textit{q}-axis priority CL}

As shown in Fig.~\ref{Fig_current limiting}(c), \textit{q}-axis priority CL keeps $i_q^{r e f}$  as constant as possible and achieves current limiting by cutting $i_d^{r e f}$, $i_q^{\text {ref※ }}$ and $i_d^{\text {ref※ }}$ are determined by~\eqref{equ:q}.

\begin{equation}
\label{equ:q}
	\begin{aligned}
		& i_q^*=\frac{i_q^{\text {ref }}}{\left|i_q^{\text {ref }}\right|} \times \min \left(\left|i_q^{\mathrm{ref}}\right|, I_{\max }\right) \\
		& i_d^*=\frac{i_d^{\text {ref }}}{\left|i_d^{\text {ref }}\right|} \times \min \left(\left|i_d^{\text {ref }}\right|, \sqrt{\left(I_{\max }\right)^2-\left(i_q^*\right)^2}\right)
	\end{aligned}
\end{equation}

\section{ Case Study of Different Current Limiting Strategies on Transient 
Stability }
The introduction of different current limiting strategies can have varying impacts on the transient stability of the system, \cite{impacts,review} pointed out that taking the \textit{q}-axis priority CL possesses a better transient stability, but none of them performed a mechanism analysis, ~\cite{liyapunuofu} adopted the method of Liyapunov's function but it is not enough to be intuitive, here we adopt the method of EPAC.

\subsection{Active Power Analysis under Different CL} 

Under the operating conditions where the inverter currents do not reach their limits, from~\cite{impacts}, the active power can be represent as:

\begin{equation}
\label{equ:pe}
P_e=\frac{E}{R_v^2+X_v^2}\left(X_v V \sin \delta+R_v(E-V \cos \delta)\right)
\end{equation}

\eqref{equ:pe} can be  simplified as:
\begin{equation}
\label{equ:pes}
P_e=\frac{E V}{X_v} \sin \delta .
\end{equation}
\subsubsection{Angle Priority}

Combined~\eqref{equ:angle} and~\eqref{equ:pes}. 
The implementation of angle priority CL results in a new active power curve.

\begin{equation}
P_e=\left\{\begin{array}{l}
\frac{E V}{X_v} \sin \delta,\left|i^{\text {ref }}\right|<I_{\max } \\
I_{\max } \cos \frac{\delta}{2}, \text { Otherwise }
\end{array}\right.
\end{equation}

\subsubsection{\textit{d}-axis Priority}
Combined~\eqref{equ:d} and~\eqref{equ:pes}. 
The implementation of \textit{d}-axis priority CL results in a new active power curve.
\begin{equation}
P_e=\left\{\begin{array}{l}
\frac{E V}{X_{\text {rev }}} \sin \delta,\left|i^{\text {ref }}\right|<I_{\max } \\
\frac{i_d^{\text {ret }}}{\left|i_d^{\text {red }}\right|} V I_{\max } \cos \delta,\left|i_d^{\text {ref }}\right| \geq I_{\max } \\
\frac{V^2}{2 X_v} \sin 2 \delta-V \sin \delta \sqrt{I_{\max }^2-\left(\frac{V}{X} \sin \delta\right)^2}, \mathrm{OW}
\end{array}\right.
\end{equation}
\subsubsection{\textit{q}-axis Priority}

Combined~\eqref{equ:q} and~\eqref{equ:pes}. 
The implementation of \textit{q}-axis priority CL results in a new active power curve.

\begin{equation}
P_e=\left\{\begin{array}{l}
\frac{E V}{X_v} \sin \delta,\left|i^{\text {ref }}\right|<I_{\max } \\
-V \sin \delta I_{\max } \frac{i_q^{\text {ref }}}{\left|i_q^{\text {ref }}\right|},\left|i_q^{\text {ref }}\right| \geq I_{\max } \\
V \cos \delta \sqrt{I_{\max }^2-\left(\frac{V}{X_v}(\cos \delta-1)\right)^2} \\
-\frac{V^2}{X_v} \sin \delta(\cos \delta-1), \text { OW }
\end{array}\right.
\end{equation}
\subsection{Transient Stability Analysis Using EPAC Method}

The Equal Proportional Area Criterion (EPAC) is commonly utilized in the transient stability analysis of traditional power systems. Given the adoption of the VSG control strategy in this context, the network exhibits characteristics akin to those of conventional power systems, thus justifying the application of the EPAC here. The EPAC informs us that, upon the occurrence of a transient disturbance in the power system, due to characteristic~\eqref{Equ:swing}, the rotor angle will gradually increase, corresponding to an area of acceleration. After the fault is cleared, the acceleration persists, causing the rotor angle to continue to rise but at a diminishing rate, corresponding to an area of deceleration. To ensure the system's transient stability, the area of acceleration must be less than the area of deceleration. It indicates that the rotor gains more energy during the acceleration process than it loses during the deceleration process, finaly resulting in instability.

As shown in Fig.~\ref{Fig_epac}, The red curve represents P-$\delta$ curve with CL, \( \delta_o \) is the fault occur time. \( \delta_c \) is the fault cleared time. According to EPAC, to keep transient stability of the VSG, the acceleration area must smaller than the deceleration area. Fig.~\ref{Fig_epac}(a)(b)(c) represents the transient stability analysis of angle priority CL, \textit{d}-axis priority CL, \textit{q}-axis priority CL, respectively. We can intuitively see that the \textit{q}-axis prioritized current limiting strategy has a significantly smallest acceleration area, thus, compared to other CL, it has better stability.

\section{Proposed Effective Current Limiting Strategy}

Based on the above EPAC method, different current limiting strategies affect the transient stability of VSG by affecting the acceleration area, here, an effective current limiting strategy is proposed to obtain a smallest acceleration area for optimal transient stability.

\subsection{Mathematical Analysis}

\begin{figure}[htbp]
    \centering
    \includegraphics[width=0.6\linewidth]{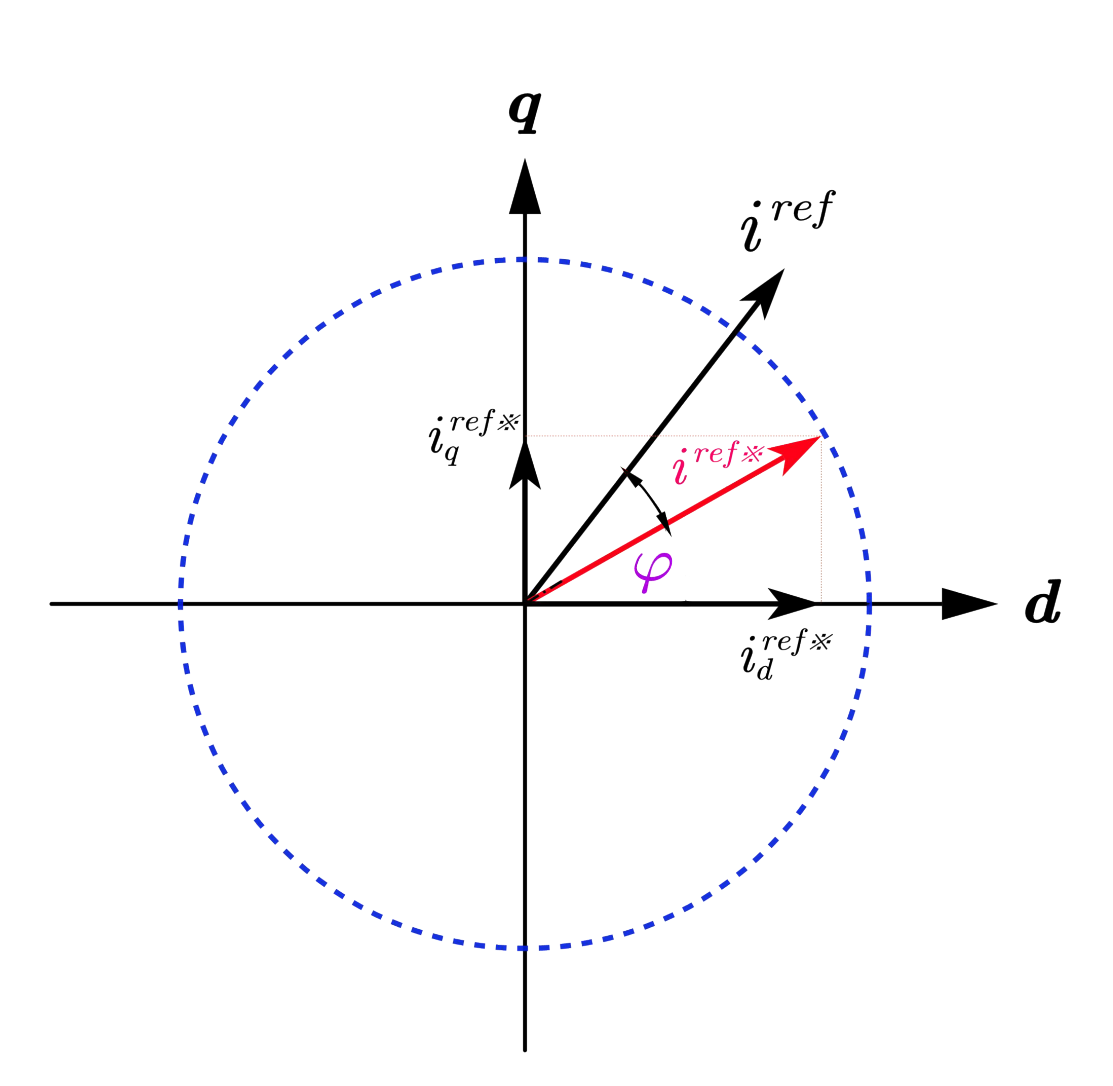}
    \caption{Adaptive current limting strategy}
    \label{fig:adaptive}
\end{figure}

As shown in Fig.~\ref{fig:adaptive}, Unlike other literature that adopts a prioritization approach, we achieve current limiting by adaptively adjusting the current angle after passing through the current limiter. 
It is easy to get \textit{d}-axis and \textit{q}-axis components of the PCC voltage.
\begin{equation}\label{equ:vd}
\begin{aligned}
&V_d=V \cos \delta
&V_q=-V \sin \delta
\end{aligned}
\end{equation}

When the current is not influenced by the current limiting strategy, it can be represented as:

\begin{equation}
I=\frac{E \angle \theta_m-V \angle \theta_g}{R_v+j X_v} .
\end{equation}

By combining with~\eqref{Equ:theta} and assuming $E \approx V$, it can be known that:
\begin{equation}
\begin{aligned}
|I| & \approx \frac{2 V}{X_v} \sin \frac{\delta}{2}   
&\angle I & \approx \theta_g+\frac{\delta}{2}
\end{aligned}
\end{equation}

When proposed adaptive current limiting control strategies are implemented, we can obtain the magnitude of the current affected by the current limiting strategies.

\begin{equation}\label{equ:i}
\begin{aligned}
& i_d^*=I_{\max } \cos \left(\frac{\delta}{2}+\varphi\right) 
& i_q^*=-I_{\max } \sin \left(\frac{\delta}{2}+\varphi\right)
\end{aligned}
\end{equation}

By substituting \eqref{equ:vd} and \eqref{equ:i} into \eqref{equ:pe}, we can derive the expression for the active power that has been influenced by the adaptive current limiting strategy.

\begin{equation}
P_e^*=V I_{\max } \cos \delta \cos \left(\frac{\delta}{2}+\varphi\right)+V I_{\max } \cos \delta \sin \left(\frac{\delta}{2}+\varphi\right)
\end{equation}

Which can be simplified as follows:

\begin{equation}
\begin{aligned}
P_e^* & =V I_{\max }\left[\cos \delta \cos \left(\frac{\delta}{2}+\varphi\right)+\sin \left(\frac{\delta}{2}+\varphi\right) \sin \delta\right] \\
& =V I_{\max } \cos \left(\varphi-\frac{\delta}{2}\right)
\end{aligned}
\end{equation}

\subsection{Transient Stability Analysis}
 To enhance the transient stability of the VSG, it is essential to minimize the acceleration area. This can be achieved by continuously adjusting the parameter $\varphi$ to ensure that  \( P_e-P_e^* \) remains at its minimum value, If precision is not critical, a simplification can be made by adjusting $\varphi$=$\frac{\delta}{2}$ to minimize \( P_e-P_e^* \), thereby obtaining a smallest acceleration area.

\begin{figure}[htbp]
    \centering
    \includegraphics[width=0.6\linewidth]{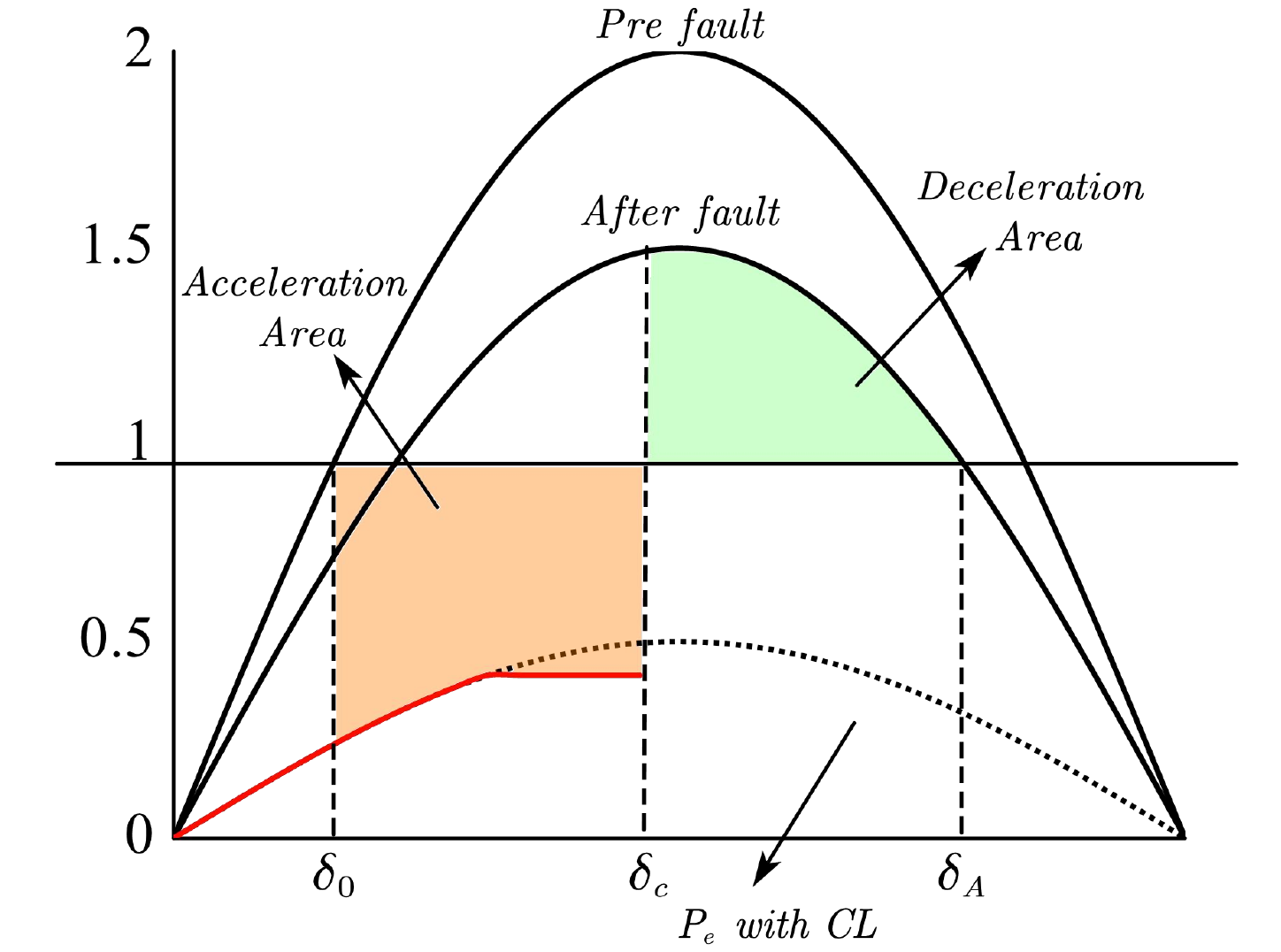}
    \caption{Transient stability analysis of proposed CL}
    \label{fig:adpative epac}
\end{figure}

Fig.~\ref{fig:adpative epac} presents a transient stability analysis under an adaptive current limiting strategy. In comparison with previous current limiting strategies, it is evident that there is a smaller area of acceleration while the area of deceleration remains constant. This implies that when the system encounters the same transient disturbances, the proposed current limiting strategy exhibits superior transient stability.

\section{Simulation results}

To validate the effective CL proposed in Section 4, the VSG system in Fig.~\ref{Fig1 topology} is implemented in MATLAB/Simulink and the data provided in Table \ref{tab:Constraint}. Given that the \textit{q}-axis priority CL has been demonstrated to be optimal than angle priority, \textit{d}-axis priority in \cite{impacts}, the subsequent comparison focuses on the proposed effective CL against the \textit{q}-axis priority CL. Faults are introduced at 0.5 seconds and cleared at 0.8s. 

\begin{table}[htbp]
	\centering
	\caption{{Parameters and constraint condtions of design.}}
	\centering \begin{tabular}{c|c|c|c}
		\toprule
		Parameters & Value & Parameters & Value  \\
		\midrule
		$P_\text {ref }$ & $1000$ & $J$ & $3$	\\
		$E_\text {ref }$ & $380V$ & $D$ & $100$	\\
		$\omega_0$& $314rad/s$ & $L_f$ & $1mH$     \\
		$V_g$	& $380V$ & $C_f$	& $50mH$	\\
		\bottomrule
	\end{tabular}%
	\label{tab:Constraint}%
\end{table}%
\begin{figure}
    \centering
    \includegraphics[width=0.9\linewidth]{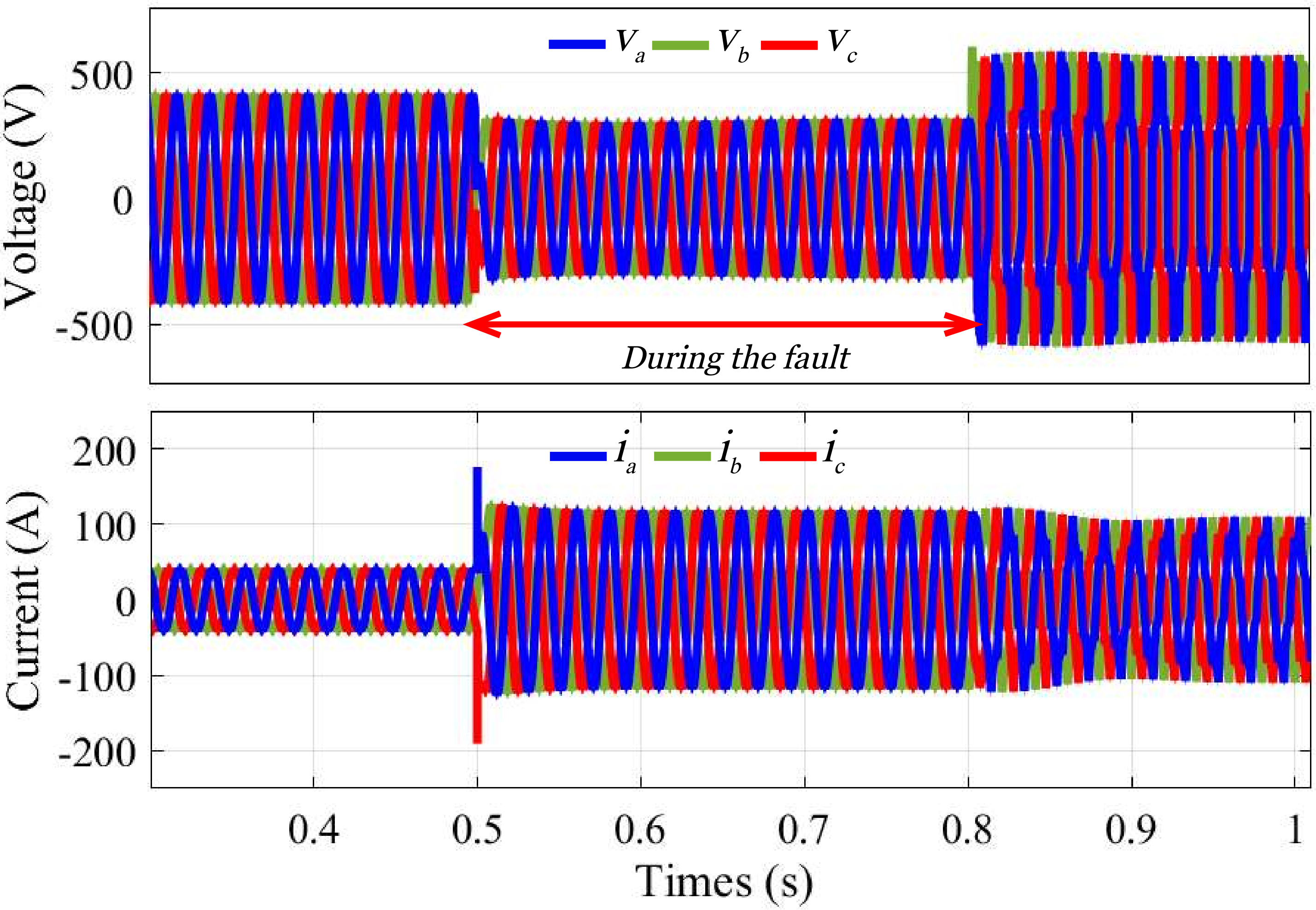}
    \caption{Voltage and current waveform under \textit{q}-axis priority CL.}
    \label{Fig_wavelfrom q}
\end{figure}

Fig.~\ref{Fig_wavelfrom q} illustrates the voltage and current waveforms for \textit{q}-axis priority CL, it can be seen in Fig.~\ref{Fig_wavelfrom q}, voltage drops at 0.5 seconds and recover at 0.8 seconds. While the current increases at 0.5 seconds, due to the current limiter, the amplitude of the current keep at 2.4pu.

\begin{figure}
    \centering
    \includegraphics[width=0.9\linewidth]{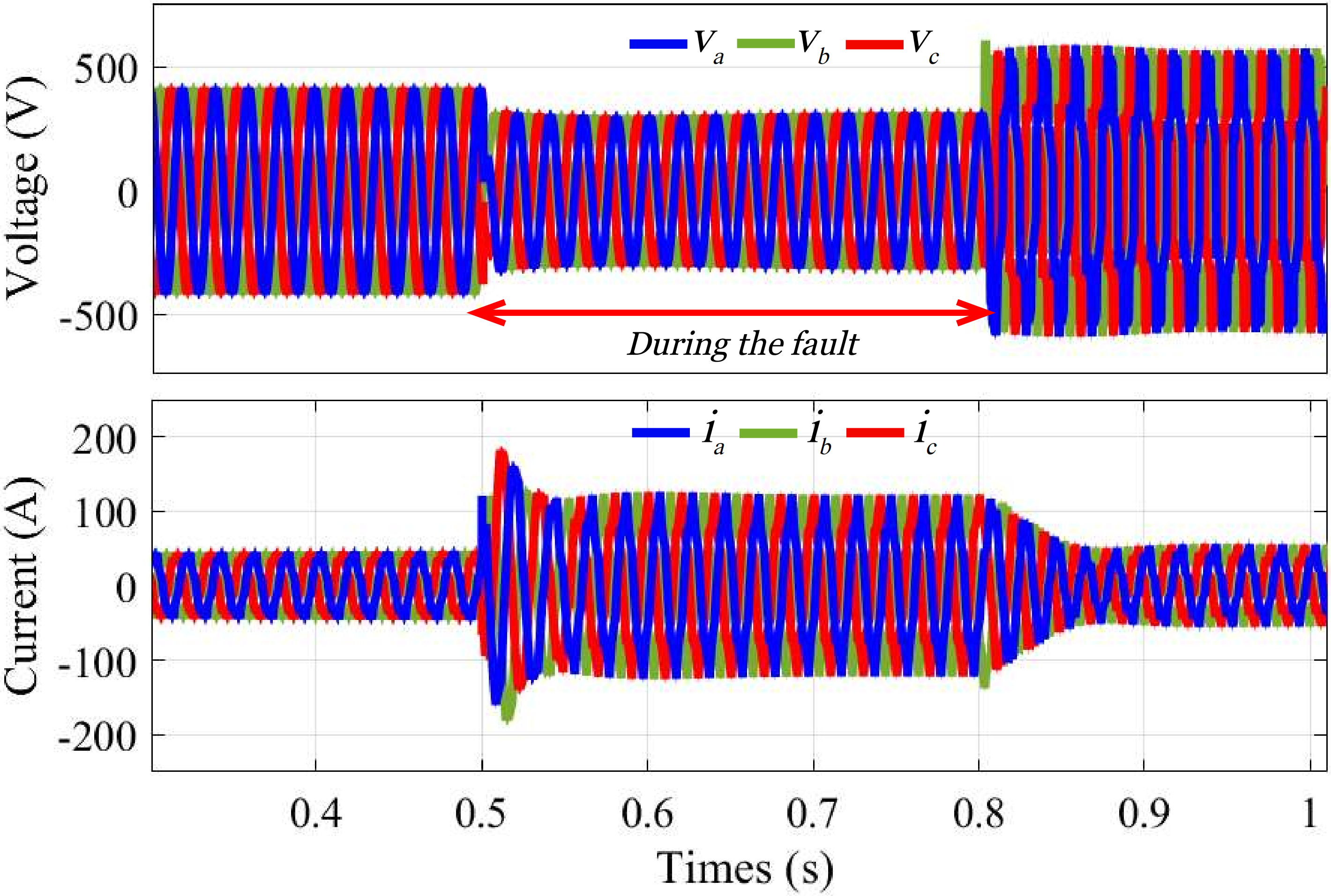}
    \caption{Voltage and current waveform under proposed effective CL.}
    \label{Fig_waveform adaptive}
\end{figure}

Fig.~\ref{Fig_waveform adaptive} illustrates the voltage and current waveforms for proposed adaptive priority CL, it can be seen in Fig.~\ref{Fig_waveform adaptive}, voltage drops at 0.5 seconds and recover at 0.8 seconds. While the current increases at 0.5 seconds, due to the current limiter, the amplitude of the current keep at 2.4pu. 
\begin{figure}
    \centering
    \includegraphics[width=0.8\linewidth]{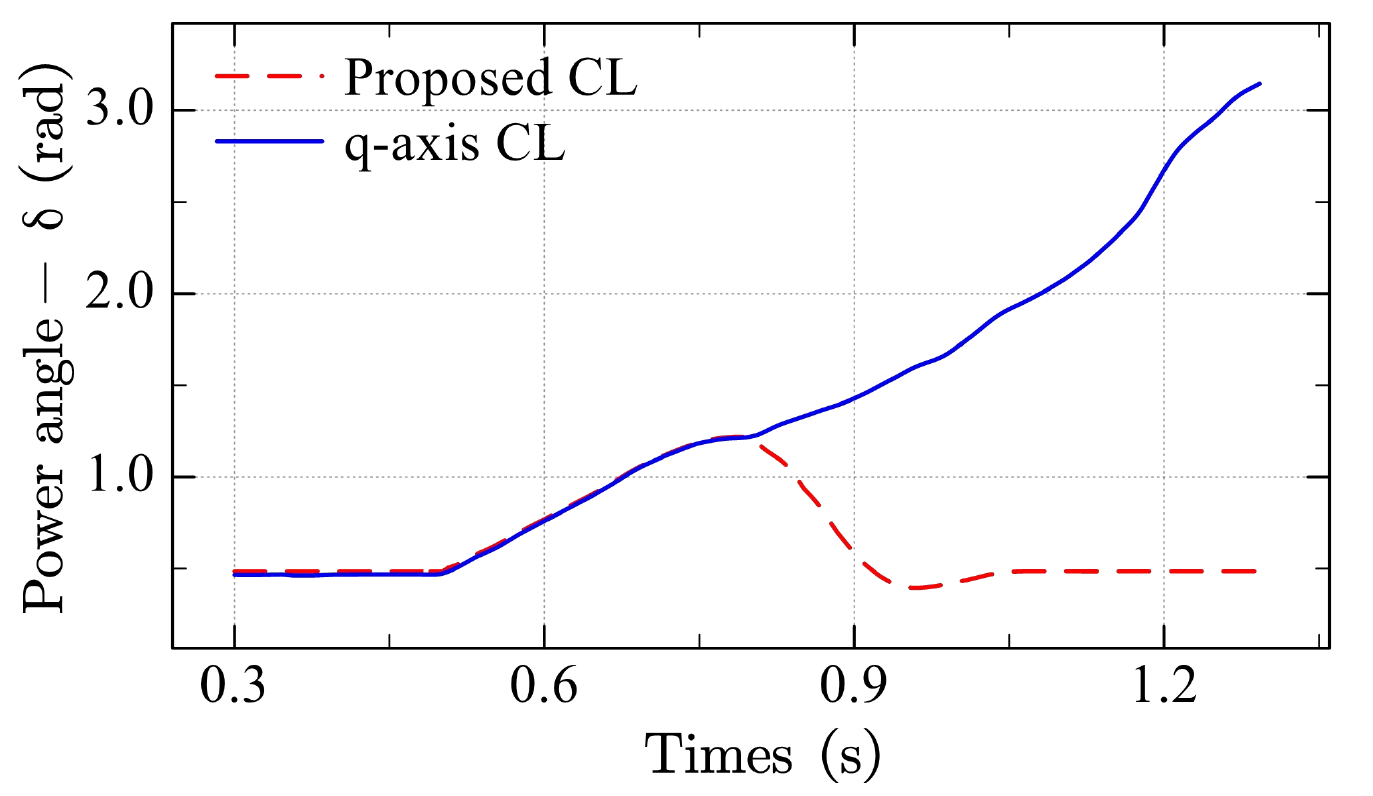}
    \caption{Power angle curve under \textit{q}-axis priority CL and proposed adaptive CL}
    \label{Fig_theta}
\end{figure}

Fig.~\ref{Fig_theta} shows the power angle of VSG under proposed adaptive priority CL, it can be found the power angle increases at 0.5 seconds. After fault cleared at 0.8 seconds, the power angle decreased, it indicates the system maintain its transient stability.
While the power angle of VSG under \textit{q}-axis priority CL, it can be found the power angle increases at 0.5 seconds. After fault cleared at 0.8 seconds, the power angle keeps increasing, it indicates the system lose its transient stability.

\section{Conclusions}
In this paper, regarding different current limiting strategies, previous studies lacked intuitive theoretical analyses. We utilize the EPAC method to provide an intuitive explanation of why different current limiting strategies exhibit varying impacts on transient stability. Additionally, to enhance the system's transient stability, an effective current limiting strategy is proposed based on this analytical approach. MATLAB/Simulink simulation results indicate that the proposed adaptive CL outperforms other strategies in terms of transient stability. 
\bibliographystyle{IEEEtran}
\bibliography{IEEEabrv,name}

\vfill

\end{CJK}
\end{document}